\documentclass[floats,floatfix,showpacs,amssymb,prd,twocolumn,superscriptaddress,nofootinbib,aps]{revtex4-1}

\usepackage{graphicx, epsf, epsfig, amssymb}
\usepackage{bm}
\usepackage{longtable}
\usepackage[usenames,dvipsnames]{xcolor}
\usepackage{color}
\usepackage[breaklinks]{hyperref}
\usepackage{amsfonts,amsmath,amssymb,mathrsfs}

\def\be{\begin{equation}}
\def\ee{\end{equation}}
\def\beq{\begin{eqnarray}}
\def\eeq{\end{eqnarray}}
\def\nn{\nonumber}

\begin{document}

\title{Black holes with massive graviton hair}

\author{Richard Brito}
\email{richard.brito@ist.utl.pt}
\affiliation{CENTRA, Departamento de F\'{\i}sica, Instituto Superior
  T\'ecnico, Universidade T\'ecnica de Lisboa - UTL, Avenida~Rovisco Pais
  1, 1049 Lisboa, Portugal}

\author{Vitor Cardoso}
\affiliation{CENTRA, Departamento de F\'{\i}sica, Instituto Superior
  T\'ecnico, Universidade T\'ecnica de Lisboa - UTL, Avenida~Rovisco Pais
  1, 1049 Lisboa, Portugal}
\affiliation{Perimeter Institute for Theoretical Physics
Waterloo, Ontario N2J 2W9, Canada
}
\affiliation{Department of Physics and Astronomy, The University of Mississippi, University, MS 38677, USA.
}

\author{Paolo Pani}
\affiliation{CENTRA, Departamento de F\'{\i}sica, Instituto Superior
  T\'ecnico, Universidade T\'ecnica de Lisboa - UTL, Avenida~Rovisco Pais
  1, 1049 Lisboa, Portugal}
\affiliation{Institute for Theory and Computation, Harvard-Smithsonian
CfA, 60 Garden Street, Cambridge, MA 02138, USA}

\date{\today}

\begin{abstract}
No-hair theorems exclude the existence of nontrivial scalar and massive vector hair outside
four-dimensional, static, asymptotically flat black-hole spacetimes. 
We show, by explicitly building nonlinear solutions, that black holes {\it can} support massive graviton hair in theories of massive gravity. 
These hairy solutions are, most likely, the generic end state of the recently discovered monopole instability of Schwarzschild black holes
in massive graviton theories.
\end{abstract}

\maketitle

\section{Introduction}
Schwarzschild black holes (BHs) stand out among all possible solutions of general relativity as the only static,
asymptotically flat, regular solution of vacuum Einstein equations. They are, in addition, stable solutions of the theory.
Perhaps surprisingly, the Schwarzschild solution also solves many other field equations, such as generic scalar-tensor theories,
$f(R)$ theories and Chern-Simons gravity (see e.g. Refs.~\cite{Psaltis:2007cw,Sotiriou:2011dz}). In fact, it is possible to show that the Schwarzschild solution is the only static,
asymptotically flat, regular solution also in the vacuum of these theories.

These uniqueness properties are in agreement with various ``no-hair'' proofs that Schwarzschild BHs cannot support
regular scalar hair, nor fields mediating the weak or the strong interaction \cite{Bekenstein:1971hc,Bekenstein:1972ky}.

The case of spin-2 hair is much less clear. It was shown by Bekenstein that BHs cannot support massive spin-2 fields in theories with generic nonminimal couplings to curvature, at least
as long as the graviton mass is sufficiently large~\cite{Bekenstein:1972ky}. 
However, as proved by Aragone and Deser~\cite{Aragone:1971kh,Aragone:1979bm}, it is impossible to couple consistently a spin-2 field with a nonlinear gravitational theory. This result does not leave much room for BHs with spin-2 hair, unless the massive tensor field is itself the mediator of the gravitational interaction, i.e. in the case of massive theories of gravity~\cite{deRham:2010ik,deRham:2010kj,Hassan:2011hr}.

Even in the case of massive gravity, recent searches for nonlinear spherically symmetric solutions~\cite{Volkov:2012wp} seem to put a rest to the possibility of finding static, asymptotically flat BH solutions endowed with spin-2 hair.

On the other hand, the nonexistence of hairy BHs in massive gravities seems at odds with the recent finding that
Schwarzschild BHs are unstable in generic theories with light massive spin-2 fields~\cite{Babichev:2013una,Brito:2013wya,Brito:2013yxa}.
The instability is due to a propagating spherically symmetric degree of freedom and it is a long-wavelength instability. It only occurs for a nonvanishing mass coupling $\mu M_S\lesssim0.438$, with $\mu$ being the graviton mass and $M_S$ the mass of the background BH (hereafter we use $G=c=\hbar=1$ units).

Interestingly, for values of $M_S$ and $\mu$ that are phenomenologically relevant, the mass coupling $\mu M_S$ is always well within the instability region. Indeed, it is natural to consider the graviton mass of the order of the Hubble constant, $\mu\sim H\sim 10^{-33}{\rm eV}$, in order to account for an effective cosmological constant (see e.g. Ref.~\cite{Hinterbichler:2011tt}). This tiny value implies that a graviton with mass $\mu\sim H$ would trigger an instability for any Schwarzschild BH with mass smaller than $10^{22} M_\odot$! 
Even if the instability timescale $\tau$ can be extremely long ($\tau\sim1.43/\mu$ in the small-mass limit~\cite{Brito:2013wya}), as a matter of principle if Schwarzschild BHs are unstable in massive gravity, they must decay to {\it something} (or not even be formed in the first place) and, unless cosmic censorship is violated, the final state should be a spherically symmetric BH. 

This apparent conundrum prompts the following question, which motivates the present study: \emph{do spherically symmetric, asymptotically flat BH solutions surrounded by a graviton cloud exist in theories with a massive graviton?}
Here, we show that such solutions do indeed exist and were not found in the thorough analysis of Ref.~\cite{Volkov:2012wp} simply because they were not searched for explicitly.

\section{Setup}
Because our purpose is merely to show that hairy BHs do exist in theories of massive gravity, we focus on a specific example
of such theories and consider the most general ghost-free Lagrangian of two interacting spin-2 fields, without matter couplings, given by~\cite{Hassan:2011zd}
\be
{\cal L}=\sqrt{|g|}\left[m_g^2R_g+m_f^2\sqrt{{f}/{g}}\, R_f-2m_v^4\, V\left(g,f\right)\right]\,. \label{biaction}
\ee
Here $R_g$ and $R_f$ are the Ricci scalars corresponding to $g_{\mu\nu}$ and $f_{\mu\nu}$, respectively; $m_g^{-2}={16\pi G}=16\pi$, $m_f^{-2}={16\pi \mathcal{G}}$ are the corresponding gravitational couplings, and $m_v$ is written in terms of $m_g$, $m_f$ and of the parameters of the potential term. The quantities $f,g$ denote the determinant of the respective metric. The potential is schematically written as
\be
\label{potential}
V\equiv\sum_{n=0}^4\,\beta_n V_n\left(\gamma\right)\,, \quad \gamma^{\mu}\,_{\nu}=\left(\sqrt{g^{-1}f}\right)^{\mu}\,_{\nu}\,,
\ee
where $\beta_n$ are real parameters and
\beq
V_0&=&1\,,\quad
V_1=[\gamma]\,,\quad
V_2=\frac{1}{2}\left([\gamma]^2-[\gamma^2]\right)\,,\nn\\
V_3&=&\frac{1}{6}\left([\gamma]^3-3[\gamma][\gamma^2]+2[\gamma^3]\right)\,,\quad
V_4=\det(\gamma)\,,
\eeq
where the square brackets denote the matrix trace.

The parameters $\beta_n$ are not all independent if flat space is to be a solution of the theory. They can be written in terms of two free parameters $\alpha_3$ and $\alpha_4$ defined as
\be
\beta_n=(-1)^{n+1}\left(\frac{1}{2}(3-n)(4-n)-(4-n)\alpha_3-\alpha_4\right)\,.
\ee
The graviton mass $\mu$ can be written in terms of the other parameters of the theory as
\be
\mu=\frac{m_v^2}{m_f} \sqrt{1+m_f^2/m_g^2}\,.
\ee

The Lagrangian~\eqref{biaction} gives rise to two sets of modified Einstein equations for $g_{\mu\nu}$ and $f_{\mu\nu}$, 
\beq
R_{\mu\nu}(g)-\frac{1}{2}g_{\mu\nu}R(g)+\frac{m_v^4}{m_g^2}\mathcal{T}^g_{\mu\nu}(\gamma)&=&0\,, \label{field_eqs1} \\
R_{\mu\nu}(f)-\frac{1}{2}f_{\mu\nu}R(f)+\frac{m_v^4}{m_f^2}\mathcal{T}^f_{\mu\nu}(\gamma)&=&0\,, \label{field_eqs2}
\eeq
where the ``graviton'' stress-energy tensors $\mathcal{T}_{\mu\nu}^g$ and $\mathcal{T}_{\mu\nu}^f$ are explicitly given by
\beq
\mathcal{T}^g_{\mu\nu}&=&\sum_{n=0}^3(-1)^n\beta_n g_{\mu\lambda}Y^{\lambda}_{\nu}(\gamma)\,,\\
\mathcal{T}^f_{\mu\nu}&=&\sum_{n=0}^3(-1)^n\beta_{4-n} f_{\mu\lambda}Y^{\lambda}_{\nu}(\gamma^{-1})\,,
\eeq
with $Y(\gamma)=\sum_{r=0}^n (-1)^r\gamma^{n-r}V_r(\gamma)$~\cite{Hassan:2011zd}.
The Bianchi identity implies the conservation conditions
\be
\nabla_g^{\mu}\mathcal{T}^g_{\mu\nu}(\gamma)=0\,,\quad \nabla_f^{\mu}\mathcal{T}^f_{\mu\nu}(\gamma)=0\,, \label{bianchi1}\\
\ee
where $\nabla_g$ and $\nabla_f$ are the covariant derivatives with respect to $g_{\mu\nu}$ and $f_{\mu\nu}$ respectively. In fact, these two conditions are not independent due to the diffeomorphism invariance of the interaction term in \eqref{biaction}, which is a general property of the ``Fierz-Pauli like'' interaction terms~\cite{Damour:2002ws}.

We consider static spherically symmetric solutions of Eqs.~\eqref{field_eqs1} and~\eqref{field_eqs2}. The most general ansatz for the metrics is given by\footnote{Note that massive graviton theories might also allow for spherically symmetric solutions whose metrics are not both diagonal in the same coordinates~\cite{Volkov:2012wp}. Since we are interested in the end state of the monopole instability found in Refs.~\cite{Babichev:2013una,Brito:2013wya}, we focus here on the ansatz~\eqref{ansatz_g}-\eqref{ansatz_f}.}
\beq
g_{\mu\nu}dx^{\mu}dx^{\nu}&=&-F^2\, dt^2 + B^{-2}\, dr^2 + R^2 d\Omega^2\,,\label{ansatz_g}\\
f_{\mu\nu}dx^{\mu}dx^{\nu}&=&-p^2\, dt^2 + b^{2}\, dr^2 + U^2 d\Omega^2\,,\label{ansatz_f}
\eeq
where $F\,,B\,,R\,,p\,,b$ and $U$ are radial functions. The gauge freedom allow us to reparametrize the radial coordinate $r$ such that $R(r)=r$. To simplify the equations we also introduce the radial function $Y(r)$ defined as $b=U'/Y$, where $'\equiv d/dr$. 

Inserting~\eqref{ansatz_g} and~\eqref{ansatz_f} into the equations of motion~\eqref{field_eqs1} and~\eqref{field_eqs2}, and using the conservation condition~\eqref{bianchi1}, we can reduce the problem to a system of three coupled first-order ordinary differential equations, which can be schematically written as (for a detailed derivation see~\cite{Volkov:2012wp})
\be
\left\{\begin{array}{l}
        B'=\mathcal{F}_1(r,B,Y,U,\mu,m_f,m_g,\alpha_3,\alpha_4)\\
	Y'=\mathcal{F}_2(r,B,Y,U,\mu,m_f,m_g,\alpha_3,\alpha_4)\\
	U'=\mathcal{F}_3(r,B,Y,U,\mu,m_f,m_g,\alpha_3,\alpha_4)
       \end{array}\right.\label{eqs_Volkov}\,.
\ee
The remaining two functions $F$ and $p$ can then be evaluated using
\beq
F^{-1}F'&=&\mathcal{F}_4(r,B,Y,U,\mu,m_f,m_g,\alpha_3,\alpha_4)\,,\label{eqs_Volkov_2a}\\
F^{-1}p&=&\mathcal{F}_5(r,B,Y,U,\mu,m_f,m_g,\alpha_3,\alpha_4)\,.\label{eqs_Volkov_2}
\eeq
The explicit form of the functions $\mathcal{F}_i$ is somewhat lengthy and not very instructive. The derivation of Eqs.~\eqref{eqs_Volkov}--\eqref{eqs_Volkov_2} and their final form is publicly available online in a {\scshape Mathematica} notebook~\cite{webpage}.

\subsection{Boundary conditions at the horizon}
Since we are interested in BH solutions, we assume the existence of an event horizon at $r_H$, where $F(r_H)=B(r_H)=0$. From the discussion in~\cite{Deffayet:2011rh,Banados:2011hk} where it is shown that for the spacetime to be nonsingular the two metrics must share the same horizon, it follows that $Y$ and $p$ must also have a simple root at $r=r_H$. On the other hand, the function $U$ can have any finite value different from zero at the horizon. For numerical purposes we then assume a power-series expansion at the horizon,
\beq
B^2&=&\sum_{n\geq 1}a_n(r-r_H)^n, \quad Y^2=\sum_{n\geq 1}b_n(r-r_H)^n,\label{BCs1}\\
U&=&u_H\,r_H+\sum_{n\geq 1}c_n(r-r_H)^n\label{BCs3}\,.
\eeq
After inserting this into the system~\eqref{eqs_Volkov}, $a_n\,,b_n\,,c_n$ all can be expressed in terms of $u_H$ and $a_1$ only, where the constant $u_H$ is arbitrary while $a_1$ is given by the solution of a quadratic equation 
\be
\mathcal{A}a_1^2+\mathcal{B}a_1+\mathcal{C}=0\,,\label{eq:a1_sol}
\ee
where $\mathcal{A}\,,\mathcal{B}\,,\mathcal{C}$ are functions of $u_H,\,r_H,\,\mu,\,m_f,\,m_g$ and $\alpha_3,\,\alpha_4$. Since there are two solutions for this equation for each choice of the parameters, there exist two different branches of solutions for the metric functions. Moreover, reality of $a_1$ requires $\mathcal{B}^2>4\mathcal{A}\mathcal{C}$, and this condition restricts the parameter space.

Inserting~\eqref{BCs1}--\eqref{BCs3} into Eqs.~\eqref{eqs_Volkov_2a} and \eqref{eqs_Volkov_2}, we find
\beq
F^2&=&q^2(r-r_H)+q^2\sum_{n\geq 2}d_n(r-r_H)^n\,,\\
p^2&=&q^2\sum_{n\geq 1}e_n(r-r_H)^n\,,
\eeq
where $d_n$ and $e_n$ can be expressed in terms of $u_H$ and of the other parameters and $q$ is an integration constant, which can be set arbitrarily and is related to the time- scaling symmetry.

Equations~\eqref{eqs_Volkov} are invariant under the following transformations:   
\beq
B(r)&\to& B(\lambda r)\,,\quad  Y(r)\to Y(\lambda r)\,,\nn\\
U(r)&\to& \frac{1}{\lambda}U(\lambda r)\,,\quad \mu\to \frac{\mu}{\lambda}\,.
\eeq
The parameter $u_H=U(r_H)/r_H$ remains invariant under the transformations above and the rescaling $r_H\to r_H/\lambda$. We use this rescaling to express all dimensionful quantities in terms of the mass of a Schwarzschild BH with horizon $r_H$, i.e. $M_S=r_H/2$.  We also consider without loss of generality $m_f=m_g$. 

Another important quantity that can be used to check the validity of the solutions is the temperature of each horizon, which can be evaluated as~\cite{Volkov:2012wp} 
\be
T=T_g\equiv\frac{q\sqrt{a_1}}{4\pi}=T_f\equiv \frac{q\sqrt{b_1 e_1}}{4\pi c_1}\,.
\ee
These two temperatures can be shown to be the same for any value of the parameters, in agreement with the discussion of Ref.~\cite{Banados:2011hk}. To evaluate the temperature we fix the constant $q$ by requiring that $F(r)\to 1$ (or, equivalently, $p(r)\to 1$) as $r\to \infty$.

\subsection{Asymptotically flat solutions}
We require the solutions to be asymptotically flat such that as $r\to \infty$, we have $B=1+\delta B$, $Y=1+\delta Y$, $U=r+\delta U$, where the variations are taken to be small. Inserting this in the field equations~\eqref{eqs_Volkov}, we obtain to first order
\beq
\delta B&=&-\frac{C_1}{2r}+\frac{C_2(1+r\mu)}{2r}e^{-r\mu}\,,\label{inf1}\\
\delta Y&=&-\frac{C_1}{2r}-\frac{C_2(1+r\mu)}{2r}e^{-r\mu}\,,\label{inf2}\\
\delta U&=&\frac{C_2(1+r\mu+r^2\mu^2)}{\mu^2 r^2}e^{-r\mu}\,,\label{inf3}
\eeq
where $C_1$ and $C_2$ are integration constants.
Finally, we can find asymptotically flat solutions numerically using a shooting method. 
\section{Results}
%
\begin{figure*}[htb]
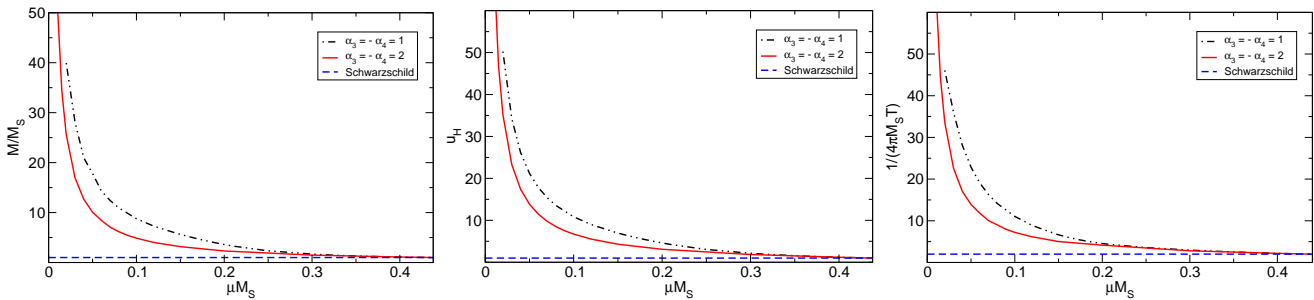

\begin{center}
\begin{tabular}{ccc}
\epsfig{file=M_vs_mu.eps,width=5.7cm,angle=0,clip=true}&
\epsfig{file=u_vs_mu.eps,width=5.7cm,angle=0,clip=true}&
\epsfig{file=T_vs_mu.eps,width=5.7cm,angle=0,clip=true}
\end{tabular}
\caption{Spacetime mass $M$ (left panel), parameter $u_H$ (center panel) and inverse of the temperature $T\equiv T_g=T_f$ (right panel) as functions of the graviton mass $\mu M_S$. Quantities are normalized by the mass of the corresponding Schwarzschild BH, $M_S$. For graviton masses close to the critical threshold $\mu M_S\sim 0.438$ the solutions merge smoothly with a Schwarzschild BH, as they should since the latter are marginally stable at this point~\cite{Babichev:2013una,Brito:2013wya}.  \label{fig:mass}}
\end{center}
\end{figure*}
%
%
%
For fixed values of $\mu,\,\alpha_3$ and $\alpha_4$, we integrate the system~\eqref{eqs_Volkov} starting from the horizon with the boundary conditions~\eqref{BCs1}--\eqref{BCs3}, towards large $r$ and find the values of the shooting parameter $u_H$ for which the solution matches the asymptotic behavior~\eqref{inf1}--\eqref{inf3}.
For each choice of $\mu,\,\alpha_3$ and $\alpha_4$, there are two branches of solutions, corresponding to the two roots of the quadratic equation~\eqref{eq:a1_sol}. In most cases only one of the branches will give an asymptotically flat solution. 

As expected, a trivial solution for any value of $\mu$, $\alpha_3$ and $\alpha_4$ is obtained when $u_H=1$, and it corresponds to the two metrics being equal and described by the Schwarzschild solution. However, we also find different solutions for which $u_H\neq1$ and that correspond to regular, asymptotically flat BHs endowed with a nontrivial massive-graviton hair. We note that such solutions were not found in Ref.~\cite{Volkov:2012wp}, most likely because the free parameter $u_H$ was not adjusted in order to enforce asymptotic flatness.

Our results are summarized in Figs.~\ref{fig:mass}--\ref{fig:func}.
The first important result is that hairy solutions exist near the threshold $\mu M_S\lesssim 0.438$ for {\it any} value of $\alpha_3,\alpha_4$. This number precisely corresponds to the critical threshold for the Gregory-Laflamme instability~\cite{Gregory:1993vy}, which was found at the linear level~\cite{Babichev:2013una,Brito:2013wya}. Solutions were expected to exist
close to this threshold and, in fact, this expectation has prompted our search at the nonlinear level.

We also find that for any value of $\alpha_3$ and $\alpha_4$, $M/M_S$, $u_H$ and $(M_S T)^{-1}$ are monotonically increasing functions of $(\mu M_S)^{-1}$ as shown in Fig.~\ref{fig:mass}. Here $M$ is the spacetime mass evaluated from the asymptotic behavior at infinity as $M=C_1/2$ [cf. Eqs.~\eqref{inf1}--\eqref{inf3}].

Above the threshold $\mu M_S\gtrsim 0.438$, the Schwarzschild solution is linearly stable. Consistent with the linear analysis, the only asymptotically flat solution that we were able to find in this region is the Schwarzschild solution, labeled by $u_H=1$ and $M=M_S$. 

\subsection{Parameter space}
The behavior at smaller $\mu M_S$ is more convoluted as it depends strongly on higher curvature terms: the nonlinear terms of the potential~\eqref{potential} become important and the solution differs substantially from the eigeinfunctions found in Ref.~\cite{Brito:2013wya} at linearized level. Nevertheless, after an extensive analysis of the full two-dimensional parameter space $(\alpha_3\,,\alpha_4)$, we obtain the following classification: 
%

(i) $\alpha_3\neq-\alpha_4 \cup \alpha_3=-\alpha_4\lesssim 1$ -- The solutions stop to exist below a cutoff $\mu_c M_S$, where the two branches of solutions near the horizon merge.

\begin{figure}[htb]
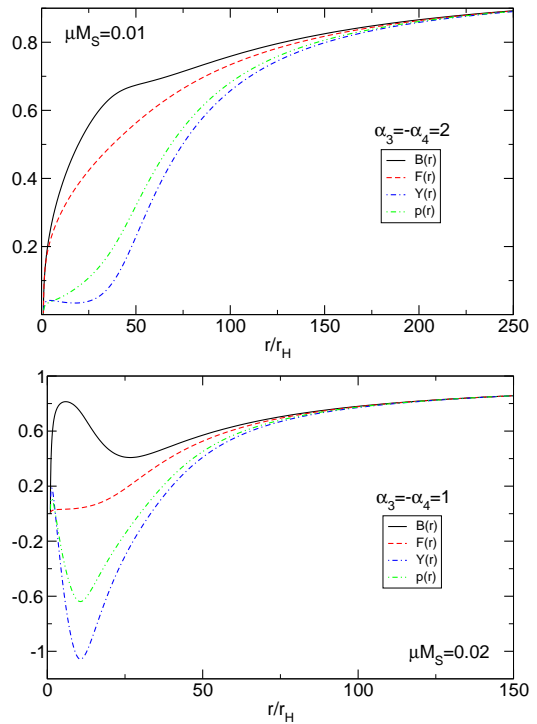

\begin{center}
\begin{tabular}{cc}
\epsfig{file=func_vs_r.eps,width=6.9cm,angle=0,clip=true}\\
\epsfig{file=func_vs_r_a3_1_a4_m1.eps,width=6.9cm,angle=0,clip=true}
\end{tabular}
\caption{Metric functions for small masses $\mu M_S$. Top panel: $\alpha_3=2\,,\alpha_4=-2$. Bottom panel: $\alpha_3=1\,,\alpha_4=-1$.\label{fig:func}}
\end{center}
\end{figure}

(ii) $1\lesssim\alpha_3=-\alpha_4\lesssim 2$ -- The solutions disappear only near $\mu M_S\sim 0.01$ and are singular at small $\mu M_S$, because some component of the metric $f_{\mu\nu}$ is vanishing where the metric $g_{\mu\nu}$ is regular (see Fig.~\ref{fig:func}). This causes the stress-energy tensor of Eq.~\eqref{field_eqs2} to become singular at these points.

%
(iii) $\alpha_3=-\alpha_4\gtrsim 2$ -- The solutions exist for arbitrarily small $\mu M_S$ and are nonsingular.
%
%
\begin{figure}[htb]
\begin{center}
\begin{tabular}{c}
\epsfig{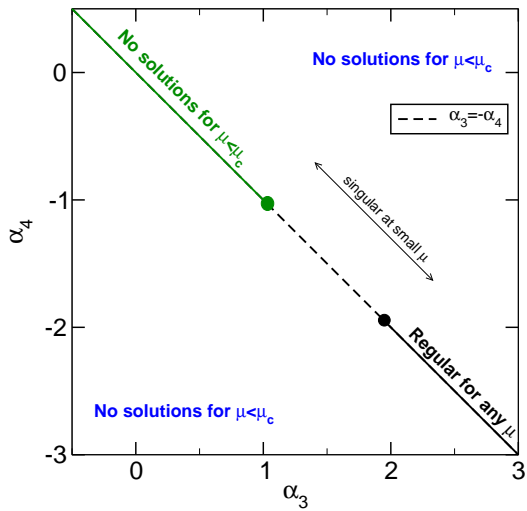}
\end{tabular}
\caption{Conjectured diagram of the parameter space for BHs with massive graviton hair in bimetric massive gravity. See main text for details.\label{fig:diagram}}
\end{center}
\end{figure}
This schematic classification of the parameter space is shown in Fig.~\ref{fig:diagram}. Although the details of this division depend on the particular model we are considering (namely, the massive bimetric theories of Ref.~\cite{Hassan:2011zd}), we believe that qualitatively similar features are likely to occur in other possible nonlinear completions of massive gravity.

It is important to emphasize that an analysis of the full parameter space is an extraordinary task. As such, it is extremely difficult to guarantee that the parameter space {\it is} divided as depicted in Fig.~\ref{fig:diagram}, as we cannot rule out certain choices of $(\alpha_3, \alpha_4)$ not belonging to the above classes. Also, the numerical integration becomes increasingly more challenging in the small-$\mu$ limit. We were able to obtain solutions for mass coupling as small as $\mu M_s\sim0.001$ and found no indication that, in the region $\alpha_3=-\alpha_4\gtrsim2$, such solutions cease to exist. However, our numerical procedure cannot guarantee that hairy BHs exist for arbitrarily small $\mu$.

\begin{figure}[htb]
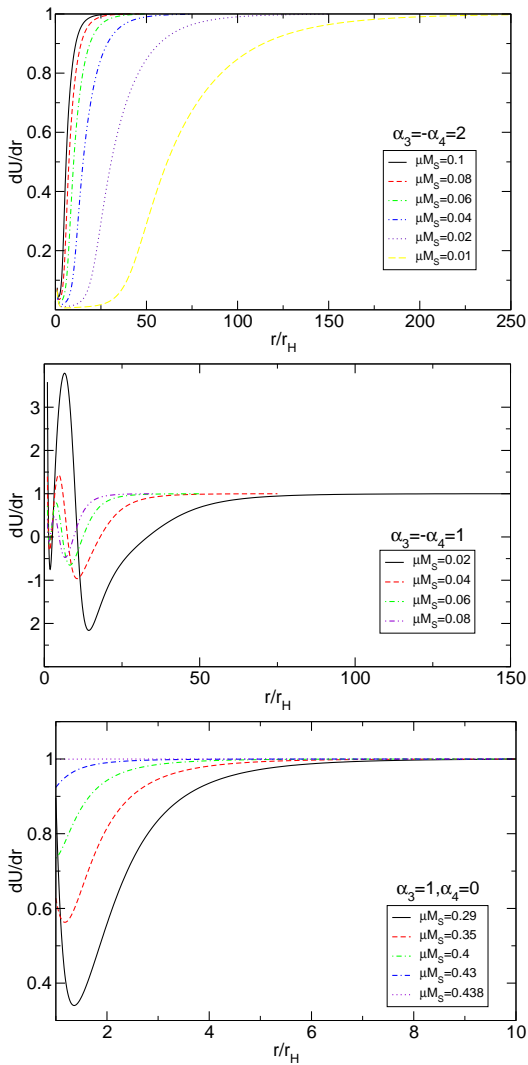

\begin{center}
\begin{tabular}{cc}
\epsfig{file=dU_vs_r.eps,width=6.9cm,angle=0,clip=true}\\
\epsfig{file=dU_vs_r_a3_1_a4_m1.eps,width=6.9cm,angle=0,clip=true}\\
\epsfig{file=dU_vs_r_a3_1_a4_0.eps,width=6.9cm,angle=0,clip=true}
\end{tabular}
\caption{The function $U'(r)$ for different values of the mass  $\mu M_S$. The behavior is similar for any value $\alpha_3$ and $\alpha_4$ near the threshold $\mu M_S\sim 0.438$ but for small $\mu M_S$ it can be very different depending on the specific values of the parameters. Top panel: $\alpha_3=2\,,\alpha_4=-2$. Middle panel: $\alpha_3=1\,,\alpha_4=-1$. Bottom panel: $\alpha_3=1\,,\alpha_4=0$.\label{fig:du}}
\end{center}
\end{figure}

The change of behavior between different regions seems to be smooth, since near the boundaries the solutions do not change drastically. For example, in the vicinity of $\alpha_3=-\alpha_4=1$ the solutions behave all in the same way.
We compare the solutions for different choices of $\alpha_3$ and $\alpha_4$ in Figs.~\ref{fig:du} and~\ref{fig:func}.

Nevertheless, the above classification seems very natural from the mathematical structure of the field equations. For instance,
the choice $\alpha_3=-\alpha_4$ corresponds to $\beta_3=0$, i.e. the higher-order term $V_3$ is absent in the potential~\eqref{potential}. Furthermore, in this case,
\beq
\beta_0&=&-6+3\alpha_3\,,\quad \beta_1=3-2\alpha_3\,,\\
\beta_2&=&-1+\alpha_3\,,\quad \beta_4=\alpha_4\,.
\eeq
Thus, the boundaries where the behavior of the solutions change qualitatively correspond to a change of sign of the parameters $\beta_n$. It is also not surprising to find that $\alpha_3=-\alpha_4=1,2$ are special points of the parameter space, because they correspond to the cases where $V_2$ and $V_0$ are, respectively, absent in the potential~\eqref{potential}.

Finally, the above picture does not hold in the limit where one of the metrics is taken to be the nondynamical Schwarzschild metric ($m_g\gg m_f$). In this case our numerical search suggests that, for any choice of $\alpha_3$ and $\alpha_4$, hairy BH solutions exist near the threshold but they do not exist for arbitrarily small $\mu M_S$. This could be explained by the fact that in the decoupling limit of massive gravity ($\mu\to 0,\,m_g\to\infty$, keeping $(\mu^2m_g)^{1/3}$ fixed) the interactions of the helicity-0 coming from the potential~\eqref{potential} can be decomposed in Galileon-like terms~\cite{deRham:2010ik}, which cannot support nontrivial configurations around a spherically symmetric BH~\cite{Hui:2012qt}.

To summarize, although it is very challenging to infer the behavior of the solutions for all choices of the parameters $\alpha_n$, we have found convincing evidence that the term $V_3$ in the potential~\eqref{potential} plays an important role as it does not allow for hairy deformations of a Schwarzschild BH in the small graviton mass limit. This term is precisely the one that gives rise to a mixing between the helicity-0 and the helicity-2 components of the massive graviton in the decoupling limit~\cite{deRham:2010ik}.

%

\section{Discussion}
As far as we are aware, the nonlinear solutions we have found are the first example of graviton-hairy BH solutions in asymptotically flat spacetime.

It is a matter of debate if the theory we considered can in fact be a viable alternative to general relativity~(see e.g.~\cite{Deser:2012qx,Burrage:2012ja,Deser:2013gpa,Deser:2013eua} and also Sec.VI of Ref.~\cite{Gabadadze:2013ria} for a recent discussion on the status of massive gravity). Nevertheless, whatever the fate of the ghost-free massive and bimetric gravities, these solutions are interesting on their own as they provide the first example of an asymptotically flat graviton-hairy BH. Furthermore, we believe that several of the properties we have presented here are likely to be found in other hairy BH solutions of any putative nonlinear theory of massive gravity.

These solutions are also natural candidates for the final state of the monopole instability recently uncovered~\cite{Babichev:2013una,Brito:2013wya,Brito:2013yxa}. The instability would presumably cause 
the Schwarzschild spacetime to evolve towards a hairy solution.
Depending on the parameters of the theory, however, different types of solutions exist in the highly nonlinear regime. This suggests that hairy, static, asymptotically flat BH solutions exist only in certain regions of the parameter space. This in turn makes nonlinear time evolutions
of Schwarzschild BHs highly desirable. It is of course possible that, in some regions of parameter space, Schwarzschild BHs
are not the preferred outcome of gravitational collapse or even that these theories do not allow for stable static BH solutions. These issues can only be addressed by performing nonlinear collapse
simulations.
 
\noindent{\bf \em Acknowledgments.}
We thank an anonymous referee for a careful reading of the manuscript.
R.B. acknowledges financial support from the FCT-IDPASC program through the grant SFRH/BD/52047/2012.
V.C. acknowledges partial financial support provided under the European Union's FP7 ERC Starting
Grant ``The dynamics of black holes: testing the limits of Einstein's theory''
grant agreement no. DyBHo--256667. 

This research was supported in part by the Perimeter Institute for Theoretical Physics. 
Research at Perimeter Institute is supported by the Government of Canada through 
Industry Canada and by the Province of Ontario through the Ministry of Economic Development 
$\&$ Innovation.
P.P acknowledges financial support provided by the European Community through
the Intra-European Marie Curie contract aStronGR-2011-298297.
This work was supported by the NRHEP 295189 FP7-PEOPLE-2011-IRSES Grant, and by
FCT-Portugal through projects CERN/FP/123593/2011.
%
\bibliography{ref}  

\end{document}